\documentclass{article}
\usepackage[utf8]{inputenc}
\usepackage{graphicx}
\graphicspath{ {./images/} }
\begin{document}
\title{A Possible Dark Matter Search Mission in Space}
\author{Nickolas Solomey\footnote{Presenter at The International Conference on Neutrinos and Dark Matter, 25-28 September 2022, Sharm El-Shaikh, Egypt; to be published in Letters in High Energy Physics.} \hspace{0.05mm} and Shrey Tripathi \\ Wichita State University, Kansas, USA}

\date{September 2022}

\maketitle

\begin{abstract}
    	Direct detection of dark matter continues to elude scientists' many attempts to see it interact, and still to this day the only way we know it is there is through observed gravitational effects. The many search experiments are at the point where the search for dark matter direct observation is limited by the solar neutrino background signal here at Earth. Past experiments typically use a large volume central detector looking for  energy materializing inside a detector volume that is not associated with any tracks of particles entering the volume through the surrounding active veto array and passive shielding. Here will be presented a new alternative method to see dark matter performing a search by changing the distance away from the Sun where the 1/r$^2$ law could be removed from the observations in a known predictable way. A Dark Matter detector on a spacecraft or built inside an asteroid might be possible. Many near Earth asteroids that can be easily reached by a spacecraft often have paths going in to the orbit of Venus and out to almost the orbit of Jupiter. These asteroids are made of ice, such as Crete, rubble piles of loosely bound boulders and pebbles, or a combination of the two. Landing on an asteroid where a space craft could melt its way under the surface for asteroids made mostly of ice or clawing its way into an asteroid could provide two advantages: shielding from cosmic and gamma rays and the ice that is melted to tunnel into the asteroid could become part of a much larger dark matter detector. Both of these advantages would allow a much larger dark matter detector than could have been brought with the spacecraft from Earth. 
\end{abstract}

\section{Introduction}
\hspace{5mm}Understanding dark matter beyond the simple observations of its gravitational effects is a major goal of the NASA Astrophysics division, the National Science Foundation and the USA Dept. of Energy Office of Science. The Earth is located 93 million miles from the Sun, and is a major source of solar neutrinos anywhere in the inner solar system. In addition to these solar neutrinos the Earth has a large number of terrestrial neutrino sources from atmospheric cosmic ray showers that create neutrinos, Geo-neutrinos from rocks, and man made reactor neutrinos. All of these neutrinos especially those from the Sun are the major limiting background for current searches of Dark Matter and Diffuse supernova neutrino sources [1], which need a lower neutrino background level to further improve dark matter search sensitivity. When searching for dark matter a dramatic decrease in the solar neutrino flux could extend the dark matter search limits for the lower mass region $<$10 GeV/c$^2$ by reducing the solar neutrino background effects and improve the galactic neutrino detector performance. This, in principle, can be done with a dark matter detector in a distant orbit around the Sun which would be an excellent new science asset permitting studies that could not be done on Earth. Changes in solar neutrino flux with distance from the Sun can further improve the performance of a Dark Matter or galactic neutrino detector sensitivity since the solar neutrino intensity is a 1/r$^2$ relation that dramatically changes by five orders of magnitude when going from the Earth to the current location of the Voyager 1, see Table 1; and this can be removed with simulations improve the search limits. 
\begin{table}[htb]
    \centering
    \begin{tabular}{|l|rl|}
        \textbf{Distance from Sun} & \textbf{Flux relati}&\hspace{-4.2mm}\textbf{ve to Earth} \\\hline\hline
        Mercury & 6&\hspace{-4.2mm}.4\\\hline
        Venus & 1&\hspace{-4.2mm}.9\\\hline
        Earth & 1&\hspace{-4.2mm}.0\\\hline
        Mars & 0&\hspace{-4.2mm}.4\\\hline
        Asteroid Belt & 0&\hspace{-4.2mm}.1\\\hline
        Jupiter & 0&\hspace{-4.2mm}.037\\\hline
        Saturn & 0&\hspace{-4.2mm}.011\\\hline
        Uranus & 0&\hspace{-4.2mm}.0027\\\hline
        Neptune & 0&\hspace{-4.2mm}.00111\\\hline
        Pluto & 0&\hspace{-4.2mm}.00064\\\hline
        Voyager 1 (2015) & 0&\hspace{-4.2mm}.00006\\\hline
    \end{tabular}
    \caption{Intensity of solar neutrinos at various distances from the sun, assuming a point source.}
    \label{tab:fluxTable}
\end{table}

\section{Dark Matter Search and Study Method}
\hspace{5mm}Dark Matter searches are soon approaching the solar and terrestrial neutrino background limits, and diffuse Supernova Neutrinos are difficult to search for due to the large solar neutrinos presence at Earth. The proposed spacecraft mission proposed here is the feasibility of a Dark Matter detector onboard a deep space probe, its science returns and its technological challenges.

\begin{figure}[htb]
\caption{Current Dark Matter experiments sensitivity for WIMP mass, solar neutrinos are responsible for the low energy WIMP mass from 0 to 10 GeV/c$^2$.}
\centering
\includegraphics[width=0.98\textwidth]{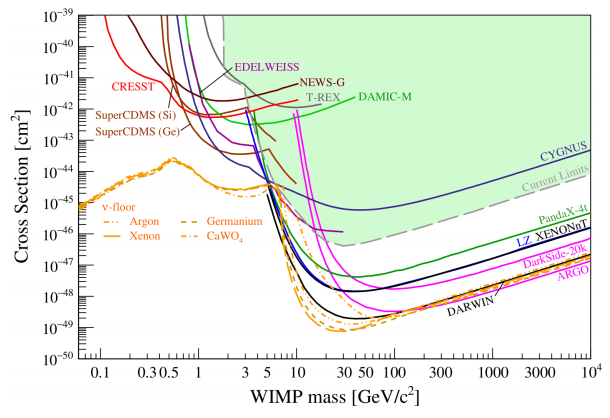}
\end{figure}

Current Dark Matter searches are approaching the “neutrino floor,” where the solar, atmospheric, reactor and Geo neutrinos are background limiting the observation of a signal (Figure 1). There is an especially high background below 10 GeV/c$^2$ mass for WIMP (Weakly Interacting Massive Particle) Dark Matter search caused by the intense solar neutrino rate at Earth. Similar to Dark Matter, the search for Diffuse Supernova Neutrinos is also limited by the background of solar and terrestrial neutrino sources. A space probe with a detector searching for these particles would very quickly be away from the terrestrial neutrino sources and by the time it reaches Saturn, it would have a hundred times fewer solar neutrinos. One obvious problem with this space probe detector method is that the size of such a detector is limited to what can be carried aloft by a rocket. Luckily, when searching for up to 10 GeV/c$^2$ mass particles, the Physics does not require as large a detector mass as the much larger mass searches for dark matter particles $>$10 GeV/c$^2$. Also there is currently a candidate Dark Matter signal right inside the solar neutrino dominated search region at 0.5 and 7 GeV/c$^2$. Another advantage over ground-based detection is that unlike on the surface of the Earth where cosmic ray showers produce a large number of secondary particles, a detector in space either in low earth orbit or on the way to planets in the outer solar system would only have to contend with the primary cosmic ray flux which is far less. However, massive shielding is needed to eliminate neutron backgrounds. Although most of these neutrinos on Earth are from air showers, i.e., secondary products of the primary cosmic rays. The only way to counter act this “noise” would be to have a fast detector that can handle the primary rates in space with several layers of active veto shielding to eliminate the charged particle cosmic ray flux. This leaves the remaining neutrino flux to be measured with a neutrino and dark-matter background rate that is sufficiently low so as to pursue a high precision dark matter search under the solar neutrino background region of Figure 1 between 1 to 10 GeV/c$^2$.

\section{Dark Matter Detector Technique}
\hspace{5mm}Currently there are 15 major ground-based, Dark Matter Search Experiments, using many different types of technologies [2]. The basic concept, see figure 2, is to have a large central detector that is sensitive to dark matter energy deposited that is recorded by photo-sensitive sensors, surrounded by an active veto and passive shielding. Although this is possible to do in a space craft, it is challenging to get a 1000 kg active detector into space and have a high performance active veto shield and passive veto [3]. Since we proposed a search and study of low mass Dark Matter then the density for Spacecraft detector only needs a smaller mass and volume to go below search limits for those of masses below the 10 GeV/c$^2$, and since this mass is where the solar and terrestrial neutrino limits are highest, we can have the greatest impact in this search region. Something else that needs to be taken into account and understood is that many of these search experiments are deep underground to reduce surface backgrounds and the best detector for deep space are those that provide fast signals so that they can be combined with fast signals from an active veto array. The need for the space craft shielding that needs to be relatively light so that it can be launched; paired with these two constraints limits the possible detector technologies to a fast scintillator signals such as those that come from cryogenic noble gases in the liquid phase, liquid scintillator or silicon detectors. 

\begin{figure}[htb]
\caption{Simple conceptual design of Dark Matter experiment for a spacecraft with inner sensitive volume using water doped with liquid scintillator, a surrounding active veto and inside a passive shield.}
\centering
\includegraphics[width=0.88\textwidth]{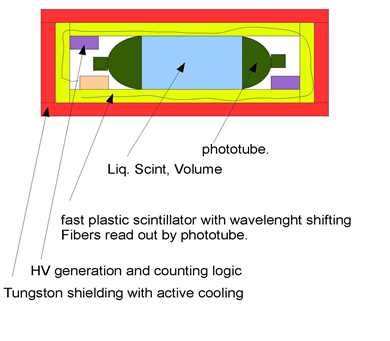}
\end{figure}

Our plan is to consider using a water based scintillator [4], using the ice of an existing asteroid to make a much larger detector. This can be done by melting a hole into the asteroid, purifying the water making a central detector volume and using a layer of melted ice and water based scintillator for an active veto layers and using the outside of the asteroid as a very large passive shield. The proposed study would also need to make sure that the intended technology used for the Dark Matter search experiment can operate for an extended period of time without the necessity of service and be kept calibrated and highly stable internally as well as with external conditions. If, for example, the external conditions do change, we will need to monitor this change so that it can be used to provide adjustments that allow us to perform a physics analysis using the known changes. All of these and similar requirements are parts of this proposed study for a Deep Space Dark Matter search experiment.

\section{Orbit for Dark Matter Asteroid}
\hspace{5mm}Although a properly built Dark Matter search experiment would have an advantage of less solar neutrino background if operated in a spacecraft or inside an asteroid further away from the Sun than Earth; there is another advantage that a spacecraft or asteroid can make use of if its radius were to dramatically change throughout an orbit. Solar neutrinos comes from the core of the Sun in a well defined shell where the fusion occurs and the intensity of these neutrinos fall off with a slightly modified 1/r$^2$ law. If dark matter has accumulated around the gravitational well formed by our Sun then these dark matter particles would have a distribution where it clumps around the Sun and falls off but it would not be the 1/r$^2$ law. A detector capable of detecting both neutrinos and dark matter, but not able to distinguish between the two, could observe, as a function of radii from the Sun, the counting rate of these combined interactions (see figure 3). Regardless of not being able to tell which is which, the 1/r$^2$ drop off for solar neutrino emissions can be removed and what remains would be the non solar neutrino emissions of our Sun. The advantage of this is that the sensitivity of such an experiment can be extremely low to dark matter (see figure 4), maybe as low as one part in a million i.e. six orders of magnitude below the solar neutrino floor in the low energy 0 to 10 GeV/c$^2$ WIMP mass region.

\begin{figure}[htb]
\caption{A simple depiction of detector counts as a function of distance from the Sun with components from Solar neutrinos, dark matter, and flat backgrounds from both stellar galactic neutrinos other than our Sun and radioactive material in the detector.}
\centering
\includegraphics[width=0.98\textwidth]{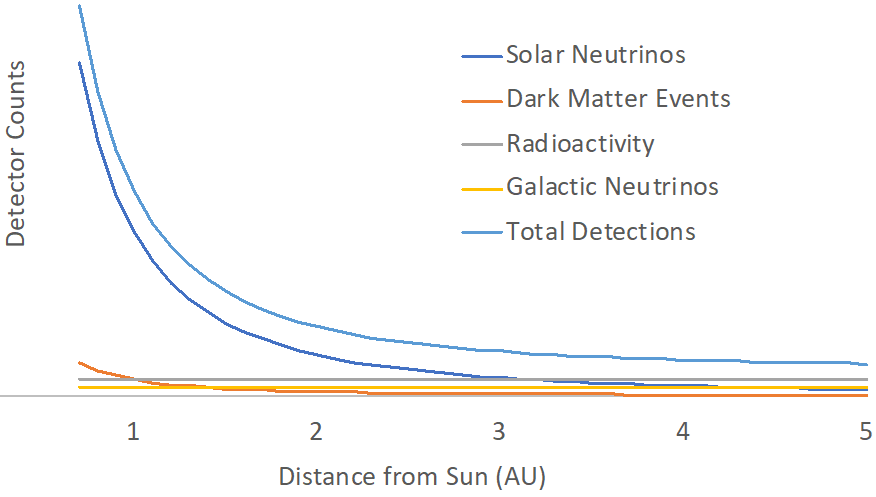}
\end{figure}
\begin{figure}[htb]
\caption{The theoretical solar neutrino floor and several curves showing the exclusion limits for a dark matter detector using the 1/r$^2$ solar neutrino fitting to remove solar neutrinos for a 25 kg spacecraft based detector (blue curve is simulations from reference [3] table 2 on page 48) and a 250 kg and 2500 kg detector the last two being a course extrapolation from the simulation; all assumed perfect veto and shielding conditions.}
\centering
\includegraphics[width=0.98\textwidth]{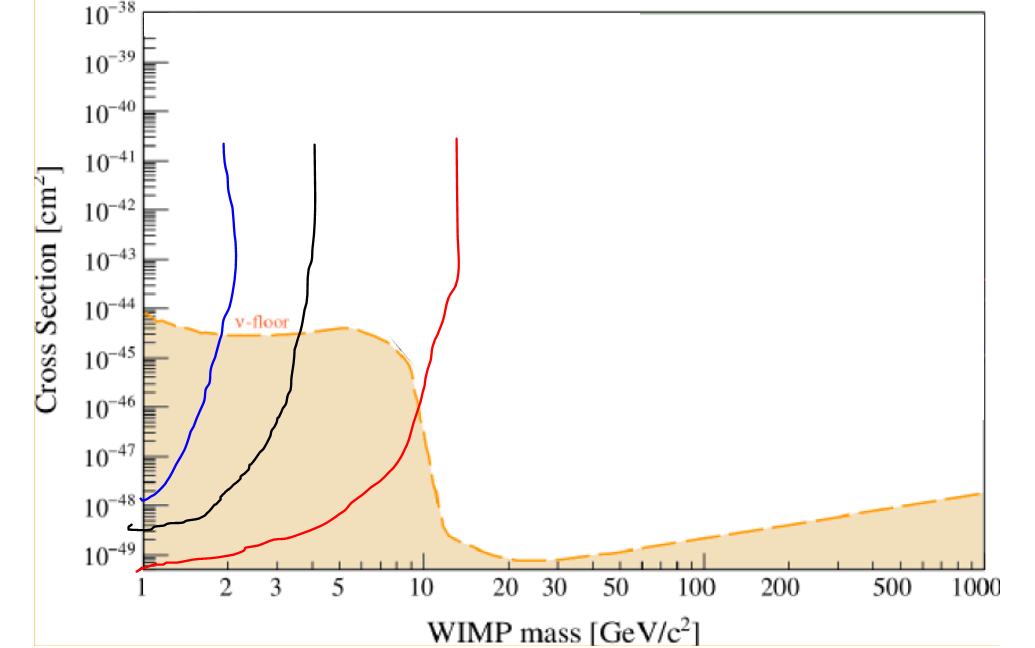}
\end{figure}

Finding an asteroid that a dark matter detector can be built inside of and has the best orbit possible is underway. This study is also surveying the types of asteroids available, which includes composition, ease of tunneling, material available to make a large dark matter search experiment, and available orbits. Although this idea is very preliminary, an asteroid such as 2018PD20 [5] where the orbit goes from Earth in towards Venus that provides lots of electrical power to melt the detector into the asteroid's center and then as it goes from the distance of Venus out to Jupiter and back in to Venus, shown in Figure 5, it can take data. This plan has many advantages and showing how the Dark Matter detector would perform on such a mission over several orbits would be an example of how the science would influence the technical aspect of any mission parameters to be investigated. There are many detector challenges that need further laboratory studies such as when water based scintillator's freeze how well does it perform and how long and how much power is needed to melt the detector into the asteroid, see the artist depiction of the fully deployed Asteroid Dark Matter detector in Figure 6.

\begin{figure}[htbp]
\caption{An Asteroid's orbit is shown that has the desired characteristics of going close to Earth as well as change distance from the Sun from the distance to Venus and out to Jupiter and back, ideal for an enhanced Dark Matter Search Experiment. [6]}
\centering
\includegraphics[width=0.98\textwidth]{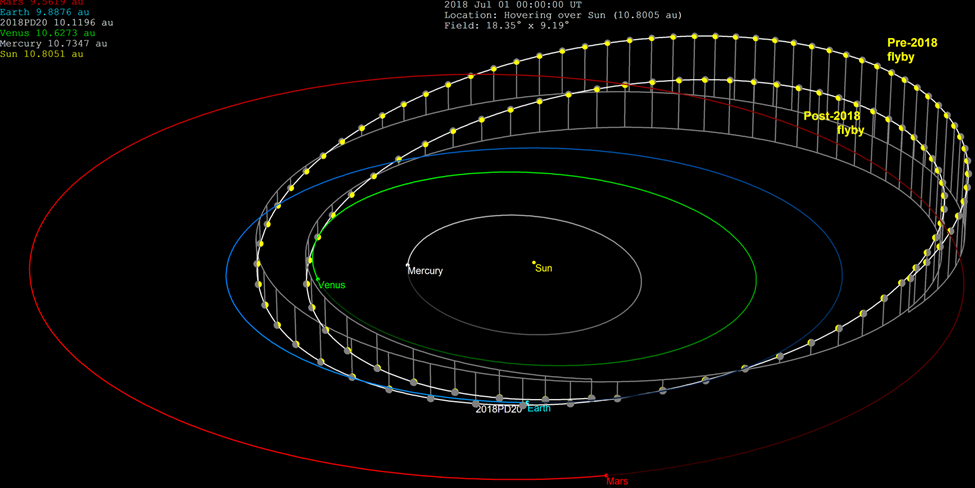}
\end{figure}
\begin{figure}[htbp]
\caption{An Artist's vision of the fully deployed Dark Matter detector inside a 25 m diameter asteroid, shown with a cutaway to reveal the detector inside.}
\centering
\includegraphics[width=0.98\textwidth]{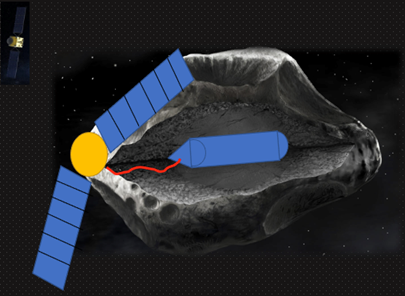}
\end{figure}

\section{Conclusion}
\hspace{5mm}Although the ideas presented here are based upon good reasoning and physics understanding, the project is in many ways very futuristic. Not only are there science studies to do so it could be understood how big a detector volume, veto array parameters and shielding inside an asteroid are necessary, but there are even more engineering challenges. Drilling a hold in the Antarctic ice shield when Ice-Cube was constructed had people present to do the work and putting down chains of photo sensors melting its way into an asteroid is similar to that effort but on an asteroid there will be no strong gravity to help guide melting the hole in the ice, mixing in a water based scintillator and doing this all with a pre-programmed robotic controls. There are also unknowns such as: does a water based scintillator still work well enough when the water freeze? A proposal for this idea has been submitted to the NASA Innovative Advanced Concept program of the NASA Space Technology Mission Directorate [6] in August 2022 and we hope to further pursue this challenging project once funded. 

\newpage
\section*{References:}

\hspace{5mm}[1] K. Eguchi et al., First Results from Kamland, PRL 90, 2003, 021802, January 17.

[2] J. Billard et al., Direct Detection of Dark Matter - APPEC Committee Report, 15 April 2021, arXiv:2104.07634.

[3] S. Tripath, Monte Carlo simulations of a space-based dark matter detector. Master’s Thesis, Wichita State University (Nickolas Solomey, advisor) April 2021.

[4] M. Yeh et al., A new water-based liquid scintillator and potential applications, NIM-A 660, p 51-56, 2011.

[5] "JPL Small-Body Database Browser: (2018 PD20)" (2018-08-12 last obs.). Jet Propulsion Laboratory. 7 September 2018.

[6] B. Sutin (JPL/Caltech) and N. Solomey (Wichita State Univ.), A Dark Matter Search Mission Inside an Asteroid, submitted to NASA STMD NIAC Phase-1 program office, August 2022.

\end{document}